\documentclass[journal]{IEEEtran}

\usepackage{tabularx}
\usepackage{amsmath,amsfonts}
\usepackage{enumitem}
\usepackage{booktabs}
\usepackage{multirow}
\usepackage{graphicx}
\usepackage{subfig}
\usepackage{tikz}
\usepackage{pifont}
\usepackage{threeparttable}
\usepackage{url}
\usepackage[most,listings]{tcolorbox}
\usepackage{makecell}
\usepackage{xspace}
\usepackage{orcidlink}
\usepackage{balance}
\usepackage{float}

\usepackage[backend=biber,style=ieee,defernumbers=true]{biblatex}
\AtBeginBibliography{\small}

\definecolor{markblue}{HTML}{1E90FF}
\newcommand{\rain}[1]{{#1}}

\newcommand{\system}{\textsc{AutoTEE}\xspace}
\newcommand{\func}[1]{\mbox{\small{\texttt{#1}}}}
\newcommand{\dquote}[1]{\textquotedblleft#1\textquotedblright}
\newcommand{\tool}[1]{\lstinline[basicstyle=\ttfamily\small, literate={-}{-}1]!#1!}

\newcommand*{\circled}[1]{\lower.7ex\hbox{\tikz\draw (0pt, 0pt)circle (.5em) node {\makebox[0.5em][c]{\small #1}};}}

\usepackage{listings}

\definecolor{colorcommentbg}{RGB}{245, 245, 245}  
\definecolor{colorcommentframe}{RGB}{0,112,155}   
\definecolor{keywordcolor}{RGB}{0, 0, 255}          
\definecolor{stringcolor}{RGB}{163, 21, 21}         
\definecolor{commentcolor}{RGB}{0, 128, 0}         

\lstdefinestyle{CodeStyle}{
    basicstyle=\ttfamily\scriptsize,
    keywordstyle=\color{keywordcolor}\bfseries,
    stringstyle=\color{stringcolor},
    commentstyle=\color{commentcolor}\itshape,
    showstringspaces=false,
    breaklines=true,
    frame=none,
    numbers=none,
    xleftmargin=1pt,
    xrightmargin=1pt,
    tabsize=1,
    aboveskip=0pt,    
    belowskip=0pt, 
}

\def\BibTeX{{\rm B\kern-.05em{\sc i\kern-.025em b}\kern-.08em
    T\kern-.1667em\lower.7ex\hbox{E}\kern-.125emX}}

\begin{document}

\title{Automated TEE Adaptation with LLMs: Identifying, Transforming, and Porting Sensitive Functions in Programs
}

\author{Ruidong Han~\textsuperscript{\orcidlink{0000-0001-6859-60057}},
        Zhou Yang~\textsuperscript{\orcidlink{0000-0001-5938-1918}},  ~\IEEEmembership{Member,~IEEE},
        Chengyan Ma~\textsuperscript{\orcidlink{0000-0001-9256-6930}},
        Ye Liu~\textsuperscript{\orcidlink{0000-0003-4562-8208}},
        Yuqing Niu~\textsuperscript{\orcidlink{0009-0003-6794-4970}}, \\
        Siqi Ma~\textsuperscript{\orcidlink{0000-0003-3479-5713}}, ~\IEEEmembership{Member,~IEEE}, 
        Debin Gao~\textsuperscript{\orcidlink{0000-0001-9412-9961}},~\IEEEmembership{Member,~IEEE},
        David Lo~\textsuperscript{\orcidlink{0000-0002-4367-7201}},~\IEEEmembership{Fellow,~IEEE}

\thanks{
Ruidong Han, Chengyan Ma, Ye Liu, Yuqing Niu, Debin Gao, and David Lo are with the School of Computing and Information Systems, Singapore Management University, Singapore (e-mail:rdhan@smu.edu.sg; chengyanma@smu.edu.sg; yeliu@smu.edu.sg; yuqingniu@smu.edu.sg; dbgao@smu.edu.sg; davidlo@smu.edu.sg).

Zhou Yang is with the School of Computing Science, University of Alberta, Canada (e-mail: zy25@ualberta.ca).

Siqi Ma is with the School of Computing and Information Technology, University of Wollongong, Australia (e-mail: siqim@uow.edu.au).

}
}

\markboth{Journal of \LaTeX\ Class Files,~Vol.~18, No.~9, September~2025}%
{Automated TEE Adaptation with LLMs: Identifying, Transforming, and Porting Sensitive Functions in Programs}
	
\maketitle

\begin{abstract}
Trusted Execution Environments (TEEs) isolate a special space within a device's memory that is not accessible to the normal world (also known as the untrusted environment), even when the device is compromised.
Therefore, developers can utilize TEEs to provide robust security guarantees for their programs, protecting sensitive operations, such as encrypted data storage, fingerprint verification, and remote attestation, from software-based attacks.
Despite the robust protections offered by TEEs, adapting existing programs to leverage such security guarantees is challenging, often requiring extensive domain knowledge and manual intervention, which makes TEEs less accessible to developers.
This motivates us to design \system, the first Large Language Model (LLM) enabled approach that can automatically identify, transform, and port functions containing sensitive operations into TEEs with minimal developer intervention.
By manually reviewing 68 repositories, we constructed a benchmark dataset consisting of 385 sensitive functions eligible for transformation, on which \system achieves an average F1 score of 0.94 on Java and 0.87 on Python.
\system effectively transforms these sensitive functions into TEE-compatible versions, achieving success rates of 91.8\% and 84.3\% for Java and Python, respectively, when using GPT-4o.
\end{abstract}


\section{Introduction}
\label{sec:intro}
\IEEEPARstart{T}{rusted} Execution Environments (TEEs) provide robust security guarantees for software programs.
TEEs create an isolated memory space that remains inaccessible to the normal world (i.e., the untrusted environment), even under a compromised operating system or hypervisor.
By executing sensitive operations (e.g., encrypted data storage, fingerprint verification, and remote attestation) within the TEE, the security and integrity of these operations can be guaranteed.
This minimizes potential security issues such as data leakage~\cite{zhang2024no,liu2022extending,zhang2020privacyscope}, unauthorized access~\cite{9664230,han2023mytee,tkdeMaLLXJSM24,wang2023tee,ait2025tee}, code tampering~\cite{ahn2020diskshield,ZhaoM19ndss}, and software-based attacks~\cite{StathakopoulouR21,zhang2019softme}.
Providing TEE support has become a standard practice for various hardware platforms across different architectures (e.g., Intel~\cite{Intel_SGX,Intel_TDX}, AMD~\cite{AMD_SEV}, ARM~\cite{ARM_TrustZone}).

Unfortunately, despite the advantages offered by TEEs and the wide support provided by mainstream platforms, adapting existing programs to leverage TEE protections is challenging.
This process requires extensive domain knowledge and manual intervention for two primary reasons:
(i) Security-critical code isolation requirements.
To minimize the trusted computing base and reduce the attack surface, developers should isolate only the security-critical code (e.g., cryptographic operations) into the TEE. However, identifying these fine-grained functions requires deep expertise in both application logic and TEE security models~\cite{zhang2024no,lind2017glamdring}.
(ii) The constrained execution environment of programs.
Most TEEs are designed to run low-level code (e.g., C or Rust) compiled directly to machine instructions, rather than high-level managed languages (e.g., Java or Python) that require dedicated runtime environments.
The overhead of securely supporting managed runtimes within the constrained environment of a TEE makes it challenging to adapt these high-level languages.
Additionally, even for low-level code, the APIs and system calls available in TEEs may differ from those in the normal world (e.g., Intel SGX does not support timers, file I/O, and multiple threads), necessitating code modifications for correct execution within TEEs.

Several methods have been proposed to facilitate TEE adoption~\cite{wuLCZWYLS24tc,miao23icse,zhang2024no,lind2017glamdring}.
One branch of research focuses on porting runtime environments for high-level languages to TEEs~\cite{wuLCZWYLS24tc,miao23icse}.
However, these approaches introduce significant overhead and may not be applicable to TEEs with small memory sizes.
The security guarantees of ported environments may not be as robust as those for native code execution, potentially exposing vulnerabilities such as memory leaks~\cite{sekeQianWZ18, kbseShahoorKYK23}.
Another research direction involves separating programs into sensitive and non-sensitive parts, and only porting the sensitive parts to TEEs~\cite{zhang2024no,lind2017glamdring}.
However, existing methods require developers to manually specify data for tracing and manually transform the program, which are complex and error-prone.

Therefore, the current practice of porting programs to TEEs is mainly a human-in-the-loop process that requires substantial manual effort~\cite{seo2017sgx, ShankerJG20sigsoft, 9276587}, motivating us to propose \system, an LLM-enabled approach that can automatically identify, partition, and transform sensitive functions in programs, and port the transformed code to TEEs.

Specifically, for a given program (in Java or Python), \system generates the Abstract Syntax Tree (AST) and extracts the leaf functions that do not invoke any other user-defined functions.
These leaf functions represent the granular units of computation within the program, which simplifies the porting process and minimizes overhead.
Then, \system prompts an LLM to review each extracted function and identify those involving sensitive operations (i.e., the use of cryptography and serialization in our paper).
Following this, \system uses LLMs to automatically transform the identified functions into their functionally equivalent native code implementations.
A successful transformation requires that the transformed code (i) compiles successfully (compilable) and (ii) preserves functional equivalence with the original code.
For functional equivalence, \system prompts the LLM to generate test inputs and employ coverage analysis tools to evaluate both line and branch coverage.
\system then provides this feedback to the LLM, enabling it to iteratively refine and improve the quality of the generated test inputs to achieve higher coverage.
Then, during the validation process, \system compares the outputs of the original function and the transformed code under the same test inputs to determine whether they remain consistent.

These steps collectively ensure that the transformed code meets both compilability and functional equivalence requirements.
To implement this goal, \system adopts the \tool{ReAct}~\cite{yaoZYDSN023} strategy to refine native code implementations, which is an iterative approach that enables LLMs to modify the generated code, validate compilability, and validate functional equivalence through external calls.
After the refinement process, \system integrates the two components through a port-based communication mechanism that includes attestation capabilities.
Specifically, for the original function, \system replaces its implementation with a remote call to the corresponding TEE-verified version.
Finally, the source program is divided into two parts: one that runs in the normal world and another that executes within the TEE.

A benchmarking dataset was manually constructed from 68 repositories in Python and Java, comprising 385 sensitive leaf functions.
Based on this benchmark, experimental results show that AutoTEE achieves an average F1-score of 0.94 on the task of sensitive function detection. 
We also evaluate the success rate of code transformation. A transformation is considered successful if, for a given set of test inputs automatically generated by an LLM, the transformed code produces outputs identical to those of the original code. 
In experiments, when applying GPT-4o for transformation, 91.8\% of Java code and 84.3\% of Python code are successfully transformed.

\vspace{0.1cm}
\noindent
\textbf{Contributions:}
\begin{itemize}[leftmargin=*]
\item We propose an automated approach, \system, that re-engineers programs for better adoption to TEEs, minimizing manual code modifications and ensuring support for high-level languages such as Java and Python.
It identifies sensitive functions within a program and transforms them into TEE-protected versions.

\item \system incorporates a robust iterative transformation and validation mechanism, utilizing compiler feedback and functional equivalence checks to guarantee the executability and functional equivalence of the TEE-adapted code.

\item We design and implement a mechanism for seamlessly integrating TEE-adapted sensitive functions into existing high-level language programs, including secure attestation and communication channels.

\end{itemize}

\section{Background and Motivation}
\label{sec:back}
\subsection{Trusted Execution Environments}\label{subsec:background:tee}
Software programs face various security risks, such as unauthorized access~\cite{9664230,han2023mytee,tkdeMaLLXJSM24,wang2023tee,ait2025tee} and data leakage~\cite{liu2022extending,zhang2020privacyscope}.
To provide stronger protection against such risks, hardware platforms and operating systems offer Trusted Execution Environments (TEEs), which are special areas in the device memory.
Logically, the memory in a device is separated into the TEE part and the normal world part.
Any computation within the TEE part is not accessible by the normal world, and thus the TEEs naturally provide security protection over many sensitive operations like algorithms, encryption keys, passwords, and personal identification information.
Developers are encouraged to run their programs involving such sensitive operations within the TEEs to minimize potential risks.
For instance, starting from version 6.0, the Android OS stores the encrypted biometric information (like fingerprints) of users and conducts the verification process within TEEs.
The normal world can only observe the verification result, lacking access to the actual verification process.
Thus, placing such sensitive operations inside TEEs can enhance the security and integrity of programs.
Currently, TEEs have become an essential part of various hardware platforms with different architectures, including Intel SGX~\cite{Intel_SGX}, Intel TDX~\cite{Intel_TDX}, AMD SEV~\cite{AMD_SEV}, ARM TrustZone~\cite{ARM_TrustZone}, and OpenTEE~\cite{OpenTEE}.

\subsection{Large Language Models \& ReAct Prompting}
Recent developments in natural language processing (NLP) have led to significant advancements in large language models (LLMs). 
These models have demonstrated remarkable improvements and have found extensive applications across various fields such as code generation~\cite{yan2023closer,zeng2022extensive}, document summarization~\cite{mastropaolo2021empirical,geng2024large}, and security analysis~\cite{wu2023effective,ren2023misuse}.
LLMs have the capability to analyze the context of a given text and make decisions based on the requirements of the task.

However, complex tasks that require logical reasoning abilities, such as code generation, bug repair, and code transformation, continue to pose challenges for LLMs. 
These tasks necessitate that LLMs decompose the complex task into simpler sub-tasks, complete these sub-tasks by providing reasoning, and subsequently make a final decision.
For instance, in code transformation tasks, LLMs must first understand the original code, then transform it into a new form, and finally validate the transformed code to ensure its equivalence with the original code.
However, LLMs primarily rely on static internal knowledge, which limits their ability to dynamically assess the quality or correctness of their transformation output through real-time interaction with external environments (e.g., compilers or runtime execution).

\tool{ReAct}~\cite{yaoZYDSN023} is a prompting strategy that enhances LLMs' reasoning capabilities by integrating external tools and actions.
It enables LLMs to engage in a \dquote{thought-action-observation} loop, interacting with external tools (e.g., databases, computation engines, search engines) to acquire additional information and dynamically validate their decisions.
It decomposes tasks into smaller sub-tasks and adapts various actions to acquire information relevant to or to complete these sub-tasks.
Therefore, LLMs can review their responses to the task and make appropriate adjustments accordingly.
Taking code transformation as an example, compiler messages can help LLMs understand whether the transformed code can be compiled, a basic requirement for successful transformation.

\subsection{Sensitive Operations}\label{subsec:background:sensitive}
In this paper, we define sensitive operations as abstract actions that directly affect program integrity and security.
Code functions that implement or directly involve these operations are classified as sensitive functions.
Specifically, we focus on cryptographic operations and serialization, following prior TEE-related research~\cite{9276587,paju2023sok}. These two categories are detailed as follows:
\begin{description}[leftmargin = 0pt]
    \item \textbf{Cryptographic operation.}
    The purpose of cryptographic operations is to ensure the security, confidentiality, and integrity of critical data.
    Previous research indicates that cryptographic operations also pose risks of leakage~\cite{han2023credential,feng2022automated} and potential attacks~\cite{dziembowski2011leakage,arikan2024tee}.
    Protective methods~\cite{alam2025tee,li2024sedcpt,niu2025trust} reveal that if these operations are conducted within TEEs, such vulnerabilities can be effectively mitigated.
    Consequently, functions that perform cryptographic operations, such as encryption, decryption, signature, verification,  hashing, and secure key management (e.g., seed generation and random number generation for cryptographic purposes), are classified as sensitive functions.
    
    \item \textbf{Serialization.}

    Serialization also poses security risks, as discussed in several studies~\cite{chen2020countermeasures,zoni2018comprehensive,graux2021preventing,chen2023tabby}.
    Insecure deserialization of untrusted data, for instance, can lead to critical vulnerabilities such as information leakage and, in extreme cases, remote code execution~\cite{sayar2023depth}.
    To address these vulnerabilities, placing the associated operations within a TEE could serve as an effective mitigation strategy.
\end{description}
While we have selected these two as representative categories, in practice, developers can adjust the prompts according to their specific requirements to identify additional sensitive functions.

\section{\system}
\label{sec:design}
\subsection{Methodology Design}\label{subsec:design:methodology}
Initially, considering the characteristics of TEEs, including resource constraints and limited support for high-level programming languages, our solution focuses on protecting the parts of the program that involve sensitive operations.
Developers often struggle to clearly identify which sections of their programs contain sensitive operations~\cite{song2021permission,cen2025ransoguard}.
To address this issue, we leverage large language models (LLMs), which are proficient in code semantic analysis~\cite{yan2023closer,wu2023effective,ren2023misuse}, to identify sensitive code sections within the program.

For the identified sensitive code, we transform it into a TEE-compatible native implementation.
This transformation process is automated using LLMs, incorporating iterative refinement with compiler checks and functional equivalence validations to ensure compilable and functional equivalence.

Subsequently, the transformed TEE-resident code is securely integrated with the original program, enabling protected execution of sensitive operations while minimizing manual intervention.

\subsection{Overview}

\begin{figure*}[ht]
    \centering
    \includegraphics[width=0.98\linewidth]{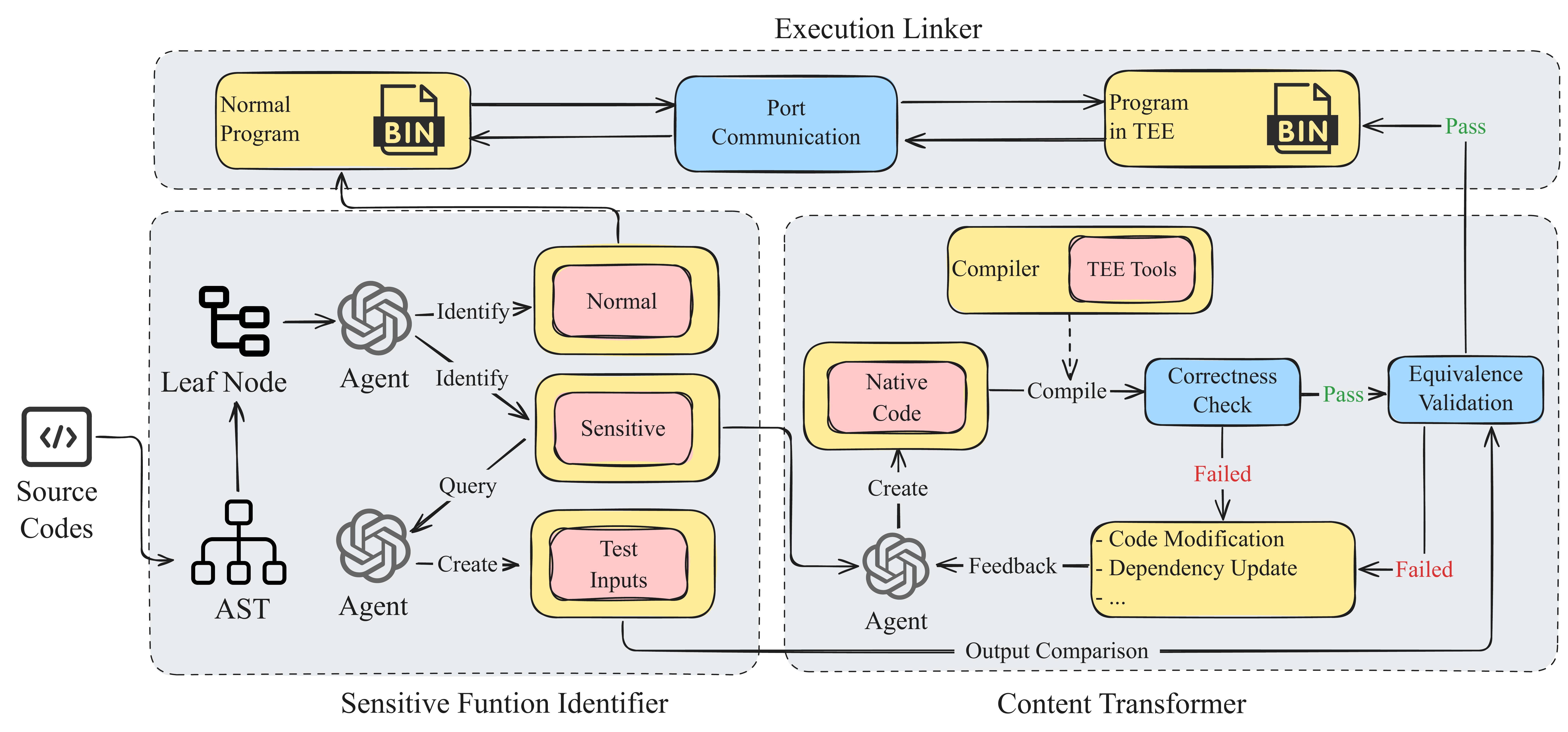}
    \caption{Workflow and three modules of \textsc{AutoTEE}. }
    \label{fig:system:overview}
\end{figure*}

Figure~\ref{fig:system:overview} illustrates the overview of our approach.
Given a program, \system first constructs the Abstract Syntax Tree and then extracts the leaf functions from the source code. 
It then employs an LLM agent to identify functions containing sensitive operations and generate test inputs.
Subsequently, \system uses a few transformation examples for the LLM to generate new native code. 
\system then checks whether the transformed code can be successfully compiled and validates its functional equivalence by comparing the outputs of the original and transformed code using the same test inputs.
If the transformation passes these checks and validations, it is considered successful.
Otherwise, \system applies the \tool{ReAct} strategy~\cite{yaoZYDSN023} to iteratively refine the generated code until it satisfies the required criteria (i.e., checks and validations). 
Once validated, the transformed code is compiled into an executable file and placed within the TEE. 
Finally, \system modifies the original program by replacing the content of the sensitive function with port-based connections that invoke the corresponding function inside the TEE.

\subsection{Sensitive Function Identifier}\label{subsec:design:identifier}
This module extracts functions from the program and identifies those that are sensitive for further transformation.

\subsubsection{Leaf Function}\label{subsub:design:leaf}
A function may involve different dependencies and additional calls.
As outlined in Section~\ref{sec:back}, TEEs operate within constrained resource environments.
Increasing the scale and complexity of code executed within a TEE can introduce performance bottlenecks and security vulnerabilities, as noted by Liu et al.~\cite{LiuDT22jss}.
Therefore, not all types of sensitive functions are suitable for porting to TEEs.
\system focuses on leaf functions, which possess the following properties:
\begin{description}[leftmargin = 0pt]
    \item \textbf{Lowest level of the program structure.}
    These functions reside at the lowest level of the program's structure and do not depend on other user-defined functions.
    They implement specific sensitive operations and serve as utilities for call chains within the project.
    Focusing on these self-contained functions avoids the complexity and potential issues associated with entire call chains.
    
    \item \textbf{Basic Arguments.}
    The code within the TEE requires interaction with the original application, which is implemented in a different programming language.
    Cross-language data interaction can introduce various compatibility issues, such as type mismatches and data structure discrepancies.
    Therefore, we restrict leaf function argument types to primitive data types (e.g., integers, floating-point numbers) and basic composite structures (e.g., arrays and strings) to ensure reliable and secure interaction.
    
    \item \textbf{Basic Library Support.}  \system can process leaf functions that invoke public libraries.
    Unlike calls to developer-defined functions, calls to standard libraries do not require \system to understand or analyze their internal implementation.
    Many commonly used libraries have equivalent implementations across different programming languages, as shown in Table~\ref{tab:design:map}.
    During porting, \system simply replaces calls to such libraries in the source language with corresponding equivalents in the TEE environment.
\end{description}

\system extracts leaf functions through a static analysis process relying on Abstract Syntax Tree (AST) processing. 
Initially, source files undergo parsing using language-specific AST tools; for instance, the \tool{ast} module is employed for Python, and \tool{JavaParser} for Java. 
\system then traverses the generated ASTs to identify function or method declarations. 
For each identified declaration, \system analyzes the function body to assess whether it invokes other user-defined functions within the project.
A function is designated as a leaf function if it does not contain calls to any such functions.

\begin{table}[htbp]
\centering
\caption{Cross-language function equivalence mappings.}
\begin{tabular}{lcc}
\toprule[1.5pt]
\textbf{Language} & \textbf{SHA-256 Hashing}                        & \textbf{Square Root}           \\
\midrule[0.8pt]
Python & hashlib.sha256()                     & math.sqrt(x)          \\
Java    & MessageDigest.getInstance("SHA-256") & Math.sqrt(x)          \\
Rust   & ring::digest::Context::new(\&SHA256)  & f64::sqrt(x)          \\
\bottomrule[1.5pt]

\end{tabular}
\label{tab:design:map}
\end{table}

\subsubsection{Sensitive Function Identification}\label{subsub:design:sensitive}
\label{subsec:sen_ident}
For the functions, \system employs an LLM to identify those that contain sensitive operations at the function level.
LLM analyzes the textual content to detect explicit indicators of security-sensitive behavior.
Therefore, sensitive functions that \system can handle should contain well-known security-related APIs (e.g., \func{hashlib.sha256}, \func{cryptography.fernet.Fernet}), sensitive keywords (e.g., encrypt, decrypt, password, token), or standard cryptographic patterns (e.g., AES-based encryption logic).
\system uses multi-round prompting, asking a series of prompts one by one.
Each prompt builds on the previous one, allowing the model to review its responses and correct prior errors.
Figure~\ref{box:sen_prompt} provides a visual overview of our prompts, with further elaboration below: green \textcolor{green!60!black}{\{Prompt\}} represents the prompt; red \textcolor{red!80!black}{\{Response\}} represents the response; orange \textcolor{orange}{\{Type\}} indicates the specific sensitive type; and cyan \textcolor{cyan}{\{Code\}} indicates the code.

\begin{figure}[ht]
  \centering
\begin{tcolorbox}[colframe=gray!50!white, 
 colback=white, 
 coltitle=black, 
 fonttitle=\bfseries, 
 left=1mm, 
 right=1mm]

\footnotesize
\textbf{- \textcolor{green!60!black}{Prompt 1:}} Does this function utilize or implement any operations related to \textcolor{orange}{[cryptography, serialization]}? Please respond with 'Yes'; otherwise, respond with 'No.' Specifically, cryptography includes \textcolor{orange}{[Encryption, ...]}; serialization includes \textcolor{orange}{[Serialization, ...]}.

\textbf{- \textcolor{red!80!black}{Response 1:}} Yes or No.

\medskip

\textbf{- \textcolor{green!60!black}{Prompt 2:}} What type of operation does this function involve?

\textbf{- \textcolor{red!80!black}{Response 2:}} [\textcolor{orange}{Encryption, Verification, ...}]. 

\textbf{- \textcolor{red!80!black}{Or:}} Sorry, the previous response is incorrect. 

\medskip

\textbf{- \textcolor{green!60!black}{Prompt 3:}} List the statements involved in [\textcolor{orange}{Encryption, Verification, ...}].

\textbf{- \textcolor{red!80!black}{Response 3:}}
[\textcolor{orange}{Encryption: \textcolor{cyan}{[code1,...]}},
\textcolor{orange}{Verification: \textcolor{cyan}{[code2, 
code3]}},
...
]

\textbf{- \textcolor{red!80!black}{Or:}} Sorry, the previous response is incorrect.

\end{tcolorbox}
\caption{Prompts for sensitive function identification.}
\label{box:sen_prompt}
\end{figure}
\begin{description}[leftmargin = 0pt]
    \item \textbf{Prompt 1:} \textbf{Code context.} 
    The initial prompt instructs the LLM to check whether a function uses cryptography or serialization operations.
    Furthermore, it specifies that the response should be either \dquote{Yes} or \dquote{No} to constrain the LLM's output and facilitate automated processing.

    \item \textbf{Prompt 2:} \textbf{Sensitive type.} 
    The second prompt identifies the type of operation involved.
    The LLM will return specific types in a list; for example, \func{[Decryption, Verification]}.
    Otherwise, the LLM will revise its previous response to \dquote{No} upon realizing that the function does not contain the corresponding operation.
    
    \item \textbf{Prompt 3:} \textbf{Statements.}
    The final prompt requests the LLM to enumerate the statements that provide evidence for sensitive judgments. 
    Similarly, the LLM may identify and correct any earlier errors in its analysis.
\end{description}
If, after these three prompts, the LLM still determines that the function contains sensitive operations, \system labels the function as \dquote{sensitive}; otherwise, it is labeled as \dquote{non-sensitive}.

\subsubsection{Test Inputs Generation}
Similar to previous approaches in code translation that use test cases to evaluate correctness at the function level~\cite{roziere2020unsupervised, chakraborty2020codit, kulal2019spoc}, \system employs test inputs for functional equivalence validation.
After identifying sensitive functions, \system instructs LLMs to generate test inputs for them. 
Inspired by recent advances in LLM-based testing~\cite{ryan2024code,alagarsamy2024a3test}, \system adopts an iterative, coverage-guided generation strategy using both line and branch coverage.
The process begins with an initial prompt that explicitly guides the LLM:

\begin{quote}
\dquote{\textit{Generate a set of diverse test inputs that invoke this function with the goal of maximizing both line and branch coverage.}}
\end{quote}
For each generated test suite, \system executes the code and measures coverage using \tool{JaCoCo} (for Java) and \tool{pytest-cov} (for Python). 
If full coverage is not achieved, the resulting coverage report is fed back to the LLM to guide the generation of additional test inputs.
This feedback loop continues until either 100\% line and branch coverage is achieved or no improvement is observed over three consecutive iterations, a stopping criterion that reflects practical limits due to LLM capability saturation or granularity in coverage reporting.

\subsection{Code Transformer}\label{subsec:design:transformer}
For the identified sensitive functions, \system transforms them into native code.
To mitigate potential vulnerabilities within the TEE, such as buffer overflows, dangling pointers, and uninitialized memory, we consider \tool{Rust}, a \emph{memory-safe} native programming language~\cite{wang2019towards,Wan3427262}, as the target for transformation.

\subsubsection{Initial Transformation}\label{subsub:design:initial}
Initially, \system prompts the LLM to analyze the code and utilizes few-shot learning techniques~\cite{snell2017prototypical} to produce an initial transformation.

\begin{description}[leftmargin = 0pt]
    \item \textbf{Prompt 1:} \textbf{What objective does this function accomplish?}
    This initial question prompts the LLM to understand the functionality of the code.

    \item [Prompt 2:] \textbf{Transform this code into Rust. 
    For example, \textcolor{cyan}{\{source code\}} to \textcolor{cyan}{\{rust code\}} with dependency \textcolor{cyan}{[A, B]}.}
    \system provides the LLM with three transformation examples, manually generated by ourselves, which the LLM can use as a reference to generate the initial transformation.

\end{description}

\subsubsection{Compilability Check and Equivalence Validation}\label{subsub:design:validation}
\label{subsubsec:check_validate}
\begin{figure}[ht]
    \centering
    \includegraphics[width=0.85\linewidth]{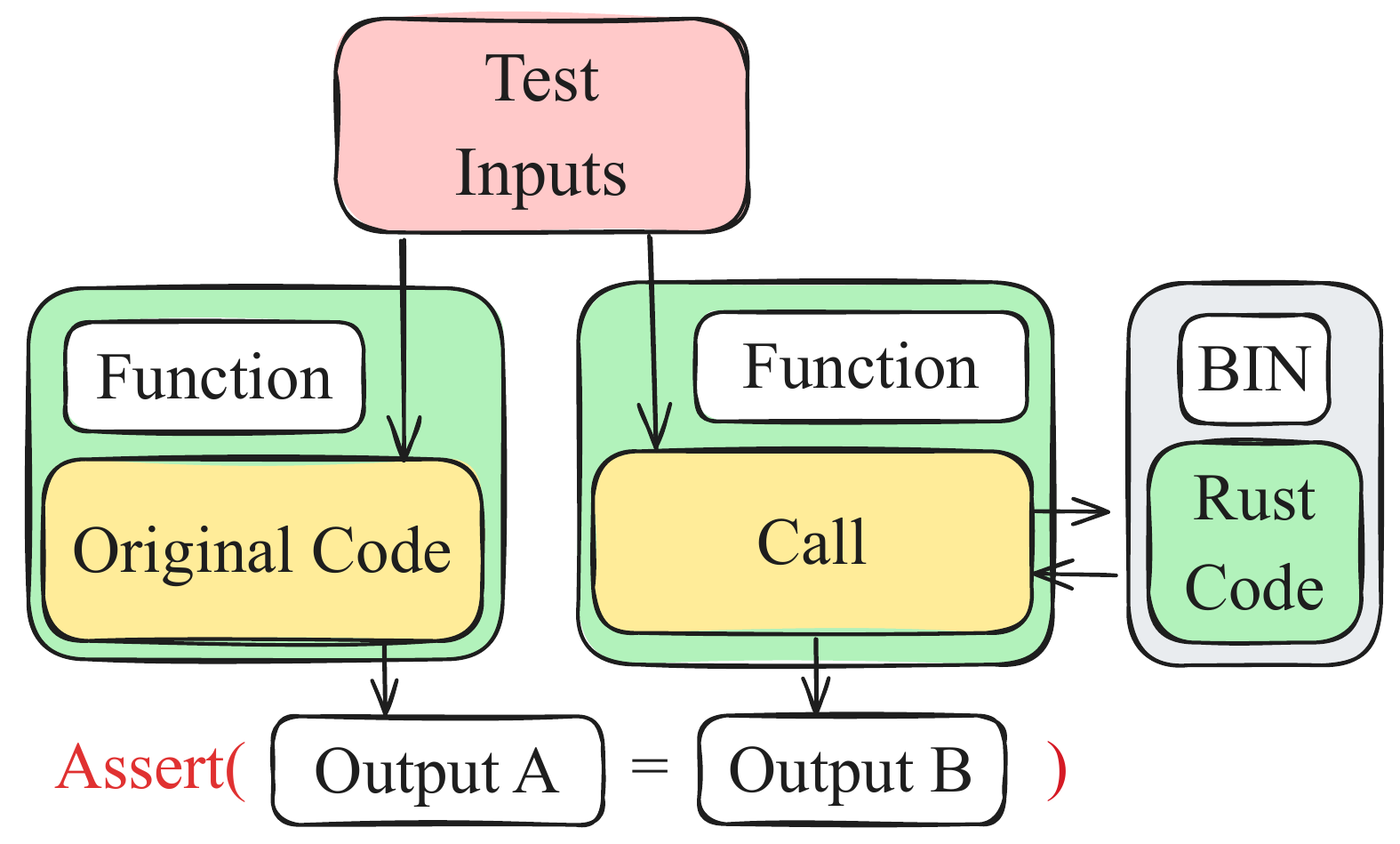}
    \caption{The Process of Validation. 
    The term BIN refers to the executable file.}
    \label{fig:system:consistent}
\end{figure}

\system first checks whether the initially transformed code is compilable (i.e., no compilation errors).
It uses \tool{Cargo} (the Rust compiler) to compile the code.
If no issues are detected, \system proceeds to validate the functional equivalence of the transformed code.
Specifically, Figure~\ref{fig:system:consistent} shows the process of this equivalence validation.
\system compiles the transformed code into an executable file.
Subsequently, \system modifies the original code function with an external call pointing to the Rust executable file.
\system then executes all test inputs to verify whether the outputs of the original and transformed code are identical.
Cryptographic operations like seed generation and key derivation inherently introduce randomness, often leading to inconsistent outputs even between functionally equivalent implementations. To establish determinism for testing or validation, we either mock the random source with predefined values or, when supported by the API, use a fixed seed.

If the code fails to compile or if the equivalence validation does not succeed, \system carries out subsequent \tool{ReAct} actions to iteratively refine the code.

\subsubsection{Iterative Refinement}
This step requires the LLM to modify the code to ensure that it successfully passes both the compilability check and the equivalence validation.
\system uses the \tool{ReAct} strategy, which is an iterative process that involves the LLM taking actions to refine the code.
Figure~\ref{box:react_prompt} illustrates example prompts utilized by \tool{ReAct}, where the blue \textcolor{blue!70!black}{{Thought/Action}} represents the thought process of \tool{ReAct}.
Following the \tool{ReAct} format, each response from the LLM will contain its \emph{Thought} along with the corresponding \emph{Action}. 
The LLM will take actions to refine the code, with each action generating feedback referred to as an observation.

\begin{figure}[ht]
  \centering
\begin{tcolorbox}[colframe=gray!50!white, 
 colback=white, 
 coltitle=black, 
 fonttitle=\bfseries,     
 left=1mm, 
 right=1mm]

\footnotesize

\textbf{- \textcolor{green!60!black}{Task Prompt:}} Implement this code using Rust with the same input type and return type \textcolor{cyan}{\{code\}}.

\textbf{- \textcolor{blue!70!black}{Thought 1:}} We need to ensure that all necessary dependencies are included in ...

\textbf{- \textcolor{blue!70!black}{Action 1:}} Dependency update (\textcolor{cyan}{[md-5, rand,...]}).

\textbf{- \textcolor{red!80!black}{Observation 1:}} Warning: use of deprecated function `base64::encode`: Use Engine::encode...

\medskip

\textbf{- \textcolor{blue!70!black}{Thought 2:}} The errors indicate issues with unused imports, missing methods, and ...; I need to address these issues by removing the unused imports ...

\textbf{- \textcolor{blue!70!black}{Action 2:}} Code modification (\textcolor{cyan}{\{rust code\}}).

\textbf{- \textcolor{red!80!black}{Observation 2:}} Error: no method named `xxx` found for struct `XXX` in the current scope...

\begin{center}
    \textit{... N time iterations ...}
\end{center}

\textbf{- \textcolor{red!80!black}{Observation N:}} Compilation successful; test input outputs are equivalent. Obtain results and exit the task.

\end{tcolorbox}
\caption{ReAct prompting to solve the code modification.}
\label{box:react_prompt}
\end{figure}

The action space defines the capacity of the LLM agent to refine the code. 
\system provides the following actions:
\begin{description}[leftmargin=0pt]

    \item \textbf{Action 1:} \textbf{Issues Search.} 
    The compiler's check process also generates an error code for specific issues (e.g., \dquote{Rust E0308}).
    The Rust official documentation includes descriptions of the index, along with example solutions.
    The LLM can leverage this action to acquire information concerning the issue and its solutions.

    \item \textbf{Action 2:} \textbf{Code Modification.}
    This action provides the LLM with an interface to modify the code and write it to the Rust file.

    \item \textbf{Action 3:} \textbf{Dependency Update.} 
    This action adds the corresponding Rust library to the Rust project, such as \func{use md5;}.
    
    \item \textbf{Action 4:} \textbf{Compilability Check.}
    This action invokes the compiler (Cargo) to check the code.
    The compiler will return the check results, indicating issues present within the code, such as the absence of packages and erroneous API calls.
    This information will serve as guidance on how to modify the code, ensuring that the transformed code is compilable.

    \item \textbf{Action 5:} \textbf{Equivalence Validation.} 
    When the code is compilable (compilability check generates no issues), the LLM will invoke this action to perform the equivalence validation.
    The process of equivalence validation is equivalent to the description provided in Section~\ref{subsubsec:check_validate}.
    If all outputs are equivalent, the LLM will receive \dquote{equivalent} feedback; 
    Otherwise, it will receive \dquote{different} feedback along with an indication of which test input is different.
\end{description}

The entire iterative process continues until the program is executable and passes equivalence validation, or until the number of iterations reaches a threshold $TH$.
$TH$ is set to 20 in our evaluation; in actual use, developers need to set it according to the specific LLM's performance characteristics.

\subsubsection{Platform Adaptation}
After several rounds of refinement, \system produces a functionally equivalent and TEE-executable code version. 
Current TEE implementations are broadly categorized into two types: VM-based (e.g., AMD SEV and Intel TDX) and process-based (e.g., Intel SGX, OpenTEE, and ARM TrustZone). 
The VM-based approach typically maintains compatibility with standard libraries, whereas the process-based approach often requires adaptations to address compatibility issues. 
As a result, for process-based TEEs, \system modifies certain API calls or system interactions accordingly.
In TEE environments, programs are launched with an empty environment, meaning no existing environment variables are available. 
When the generated code relies on environment variables, \system reads these values during preprocessing and hardcodes them directly into the source code.
Process-based TEEs do not support CPU hardware timers. 
In such cases, if the code uses CPU timer functionality, \system replaces it with a system-level timer API.

\subsection{Execution Linker}\label{subsec:design:linker}
When a sensitive function is transformed successfully, \system links the original program to its corresponding implementation in Rust.
Since these two components are in different languages and operate in distinct environments, i.e., one executing in the normal world and the other within a TEE, \system implements their interaction using a secure, port-based communication mechanism.

\begin{figure}[htb]
\centering
\includegraphics[width=0.98\linewidth]{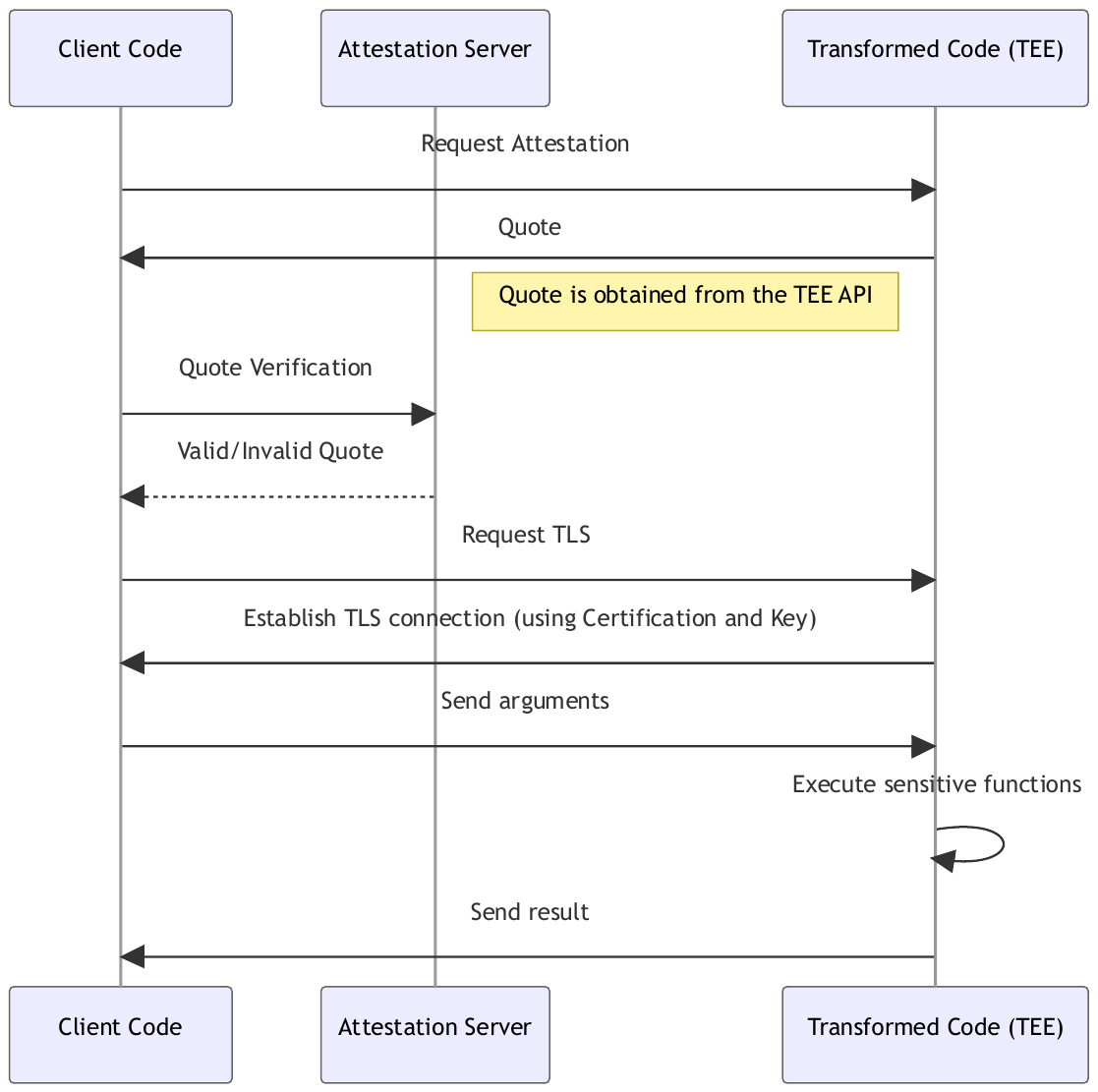}
\caption{Secure interaction between source code and TEE code.}
\label{fig:system:linker}
\end{figure}

Both components require code modifications to support a secure interaction process. 
Figure~\ref{fig:system:linker} illustrates the detailed interaction logic between the client code and the transformed code, which is defined as follows:
\begin{description}[leftmargin=0pt]
\item \textbf{Client Code.} New code logic replaces the original code, implementing TEE calls.
\item \textbf{Transformed Code.} A functionally equivalent version of the original code, generated for execution within the TEE.
\item \textbf{Attestation Server.} An external component used to verify the authenticity and integrity of the transformed code. 
These servers are typically provided by TEE SDK suites and expose open APIs for verification, such as Intel Attestation Services for SGX and AMD Key Distribution Service for AMD platforms.
\end{description}

The workflow between the client code and the transformed code is as follows:
\begin{enumerate}[leftmargin=*]
\item The client code requests the transformed code to perform attestation and return a TEE-generated report (quote).
\item The transformed code forwards the quote to the attestation server, which confirms its legitimacy and integrity.
\item Upon successful verification, the client code initiates the establishment of a secure TLS channel with the transformed code.
\item After the TLS connection is established, the client code sends encrypted function arguments to the transformed code, which decrypts and executes them within the TEE.
\item Upon completion, the transformed code sends it back to the client code over the secure TLS connection.
\end{enumerate}
After the corresponding logical modifications are made to the program's code, the transformed code will be compiled into a binary and signed using the TEE SDK.

\section{Evaluation}
\label{sec:eval}
We evaluate the performance of \system by answering the following research questions:

\begin{itemize}[leftmargin=*]
\item \textbf{RQ 1: Identification Accuracy.} Can \system accurately identify sensitive functions within a program?

\item \textbf{RQ 2: Transformation.} Can \system ensure that sensitive functions are correctly transformed, achieving the equivalent functionality as the original code?

\item \textbf{RQ 3: Resource Consumption.} What is the resource consumption in code transformations?

\end{itemize}
\subsection{Experiment Preparation}\label{subsec:eva:prep}
\label{subsec:eva:setup}
\begin{description}[leftmargin = 0pt]
\item \textbf{Experiment Environment.}
We ran \system on a server equipped with an AMD EPYC 7763 processor and an NVIDIA RTX 6000 Ada graphics card.
We chose four LLMs: GPT-4o~\cite{openai2024gpt4o}, Qwen2.5:32B~\cite{abs240912186}, DeepSeek-v3~\cite{liu2024deepseek}, and LLaMA3.1:8B~\cite{abs230709288} in our experiments, subsequently referred to as GPT-4o, Qwen2.5, DeepSeek, and LLaMA3.1, respectively.
Qwen2.5 and LLaMA3.1 are deployed locally on our server.
For TEE platforms, we selected Intel SGX (Intel i7-9700 processor).
Our evaluation includes both locally deployed models and commercial models accessed through an application programming interface. 
The objective of this mixed approach is to benchmark a diverse range of models, not to endorse the use of external services for workflows involving sensitive data.

\item \textbf{Dataset Construction.}
To the best of our knowledge, no existing dataset of functions to be protected under TEEs is available.
Thus, we manually constructed a benchmark for the task addressed in our paper.
To identify repositories that potentially contain sensitive operations, we conducted a keyword search on GitHub, targeting repository names and descriptions containing one or more of these terms: \emph{password}, \emph{credential}, \emph{seal}, \emph{crypto}, \emph{token}, \emph{serialization}, and \emph{cryptography}.
After identifying these projects, we cloned the repositories locally.
To ensure the quality of collected projects, we only kept projects with more than 50 stars, a commonly used threshold to exclude toy projects~\cite{han2023credential, feng2022automated}.
Finally, we obtained 38 Java repositories and 30 Python repositories.
By parsing these projects, we obtained 7,214 and 3,770 Java and Python functions, respectively.
We retained the leaf functions and conducted a manual review involving three authors with backgrounds in cryptography or software engineering. 
Our ground truth labels were established by human annotators using the same criteria: (1) it directly performs security-sensitive operations (e.g., encryption, authentication, key generation); (2) it explicitly invokes known cryptographic or deserialization APIs (such as hashlib.sha256 or cryptography.fernet.Fernet); or (3) it contains recognizable cryptographic patterns (e.g., AES-based encryption logic).
Only the functions that were classified as sensitive by all three reviewers were kept for further analysis.
From examples marked as sensitive, we randomly sampled two cryptography-related functions (e.g., AES encryption, RSA key generation) and one serialization function as examples that we used as \dquote{few-shot} learning instances.
The remaining sensitive functions were used in the subsequent experiments, including in 232 sensitive functions for Java and 153 for Python.
In terms of lines of code (LoC), a commonly used proxy for implementation scale in empirical software engineering research~\cite{ziegler2022productivity, chen2004model}, the average LoC is 18.2 for Java functions (range: 3–95) and 15.6 for Python functions (range: 3–83).
Cryptographic operations and serialization typically follow established usage patterns~\cite{acar2017comparing, ami2022crypto} (e.g., for encryption: initializing a key, using the key for encryption or decryption, and outputting the result), so the logic tends not to be highly complex and the LoC is generally not very large.
Additionally, to construct our dataset, we randomly added 241 Java and 166 Python normal functions from the remaining code (i.e., normal functions).

\item \textbf{Few-shot Example.}
From examples marked as sensitive, we sampled two cryptography-related functions (e.g., AES encryption, RSA key generation) and one serialization function as examples that we use as \dquote{few-shot} learning instances.
For each chosen function, one author manually implemented a Rust transformation, focusing on preserving semantic equivalence while re‑expressing syntactic structures.
The resulting transformed examples were then reviewed by the other two authors to evaluate logical equivalence.
Any discrepancies in equivalence identified during this review were resolved by manually refining the transformations.

\end{description}

\subsection{RQ1: Identification Accuracy}
\label{subsec:eva:accuracy}
\subsubsection{Identification Result}
In this experiment, we applied \system to the dataset to evaluate its effectiveness in sensitive function identification.
We repeated each experiment 10 times for each model on our benchmark and reported the average results in Table~\ref{tab:eva:detection}. 
The evaluation metrics include average precision, recall, and F1-score. 
On both datasets, LLaMA3.1 achieved the lowest F1-scores, with 0.906 on Java and 0.861 on Python, although its precision on Python was higher than that of DeepSeek and Qwen2.5. 
On the Java dataset, GPT-4o achieves the best performance across all three metrics. 
On the Python dataset, GPT-4o leads in precision and F1-score, while DeepSeek and Qwen2.5 achieve higher recall scores. Specifically, GPT-4o attains the highest average F1-score, with 0.958 on Java and 0.883 on Python, followed by DeepSeek, with 0.944 on Java and 0.881 on Python, and Qwen2.5, with 0.942 on Java and 0.872 on Python. To assess whether the observed performance differences are statistically significant, we applied the \emph{Mann-Whitney U Test}~\cite{mann1947test} to compare the F1-scores of GPT-4o with those of DeepSeek, Qwen2.5 and LLaMA3.1. For each comparison, the test was conducted on 10 F1-scores obtained from the 10 repeated experimental runs. 
On the Java dataset, GPT-4o significantly outperforms DeepSeek, Qwen2.5 and LLaMA3.1 at the 99\% confidence level ($p<0.01$). 
On the Python dataset, GPT-4o performs significantly better than Qwen2.5 and LLaMA3.1 at the 99\% confidence level ($p<0.01$), while the difference compared to DeepSeek is not statistically significant ($p>0.05$). Overall, GPT-4o demonstrates the strongest performance in sensitive function detection. DeepSeek serves as a viable secondary alternative.

\begin{table}[htb]
\centering
\caption{Performance on sensitive function detection (average). }
\label{tab:eva:detection}
\resizebox{\linewidth}{!}{
\begin{tabular}{lccccccc}
\toprule[1.5pt]
\multirow{2}{*}{Model} 
& \multicolumn{3}{c}{\textbf{Java}} 
& \multicolumn{3}{c}{\textbf{Python}} \\
\cmidrule(lr){2-4} \cmidrule(lr){5-7}
& \textbf{Precision} & \textbf{Recall} & \textbf{F1} 
& \textbf{Precision} & \textbf{Recall} & \textbf{F1} \\
\midrule[0.8pt]
GPT-4o & 94.8\%  & 96.8\%  & 0.958 & 88.1\%  & 88.5\%   & 0.883 \\
DeepSeek & 92.2\%  & 96.7\%  & 0.944 & 84.9\%  & 91.5\%   & 0.881 \\
Qwen2.5 & 94.8\%  & 93.7\%  & 0.942 & 83.8\%  & 90.8\%   & 0.872 \\
LLaMA3.1 & 90.6\%  & 90.5\%  & 0.906 & 85.4\%  & 86.9\%   & 0.861 \\
\bottomrule[1.5pt]
\end{tabular}}
\end{table}

For these failure cases, we conducted a manual examination to analyze their characteristics.
Functions that contain multiple shift operations or array manipulations may be incorrectly identified as sensitive operations because bit shifting is a commonly used operation in cryptography for security handling.
Additionally, if a function involves operations related to a dictionary with key-value pairs, the term \dquote{key} may be improperly identified as referring to cryptographic operations.
The performance gap between Java and Python in sensitive code identification can be attributed to inherent differences in language design and coding practices. 
Specifically, research has shown that statically-typed languages like Java facilitate automated code analysis due to their explicit type usage and structured APIs~\cite{ray2014large, allamanis2018survey}. 
In contrast, Python's dynamic typing, flexible syntax, and idiomatic patterns such as runtime attribute access may obscure the intent or presence of sensitive operations, making it more challenging for LLMs to detect these code segments reliably.

\subsubsection{Prompt Strategy Influence}\label{subsub:eva:prompt}
In addition, we evaluated the impact of three different prompt configurations on the detection performance.
Using Qwen2.5 as an example, we conducted another detection under three scenarios: using only prompt 1, a combination of prompts 1 and 2, and the full set including prompts 1, 2, and 3 (the setup used in the previous experiment).
The results are summarized in Table~\ref{tab:eva:detection_step}.
As shown in the table, the combined prompt configuration achieves higher accuracy compared to using a single prompt.
For Java, the F1 score increased from 0.868 to 0.909 (P1\&2) and further to 0.931 (P1\&2\&3). Similarly, Python's F1 improved from 0.791 (P1) to 0.868 (P1\&2\&3).
However, generating longer outputs inherently consumes more time and resources due to the nature of LLMs.
Our incremental prompting strategy addresses this by applying increasingly detailed prompts in stages, first filtering likely candidates with a concise prompt (Prompt 1), and then refining the results with additional prompts only when necessary.
This staged approach effectively reduces unnecessary token consumption while preserving detection accuracy.

\begin{table}[htb]
    \centering
    \caption{Identification result with different policy.}
    \label{tab:eva:detection_step}
    \begin{threeparttable}
    \begin{tabular}{@{}lccccccc@{}}
        \toprule[1.5pt]
        \multirow{2}{*}{\textbf{Qwen2.5}}       & \multicolumn{3}{c}{\textbf{Java}} & \multicolumn{3}{c}{\textbf{Python}} & \multirow{2}{*}{\textbf{R.T.}}
        \\ \cmidrule(lr){2-4} \cmidrule(lr){5-7}
                    & \textbf{P} & \textbf{R}  & \textbf{F1}   & \textbf{P} & \textbf{R}  & \textbf{F1}      \\
                    \midrule[0.8pt]
        P1    & 87.1\%   & 86.6\%  & 0.868 & 75.6\% &  83.1\%  & 0.791  & 0.3s \\
        P1\&2    &   90.9\% &  90.9\% & 0.909  &    	  79.0\%  & 86.3\% & 0.825 & 3.1s\\
        P1\&2\&3    & 93.5\%   & 92.7\%  & 0.931  & 83.8\%  & 90.9\%   & 0.868 & 7.7s \\
        \bottomrule[1.5pt]
    \end{tabular}
    \begin{tablenotes}
    \small
    \item [*] PX denotes Prompt-X; \textbf{P} denotes precision; \textbf{R} denotes recall; \textbf{F1} denotes F1-score; \textbf{R.T.} denotes the average response time in seconds.
    \end{tablenotes}
    \end{threeparttable}
\end{table}

\subsection{RQ2: Transformation}\label{subsec:eva:transformation}
This experiment uses a dataset of 232 Java functions and 153 Python functions, all of which were manually labeled as sensitive. 
By using all sensitive functions as the input for RQ2, we ensure that the evaluation of \system’s transformation effectiveness is not confounded by identification errors from RQ1.
For each of these functions, \system transforms the code into Rust.
We validated the equivalence after transformation, analyzed the reasons for failures, and compared \system with other prompting methods.

\subsubsection{Metric}\label{subsub:eva:metric}
Before launching the experiment, we established several metrics for measurement.

\begin{itemize}[leftmargin = *]
    \item \textbf{\#Sensitive:} Refers to the number of cases marked as sensitive, specifically those that require transformation.

    \item \textbf{\#Direct:} Refers to the number of cases that directly achieve executability and pass the equivalence validation after an \emph{initial transformation}, with no need for activating \emph{iterative refinement}.

    \item \textbf{\#Equivalent:} Refers to the number of cases for which functional consistency with the original code is confirmed following either an \emph{initial transformation} or \emph{iterative refinement}. This count encompasses cases categorized as \#Direct.

    \item \textbf{Avg. Iter.:} Refers to the average number of iterations completed during successful transformations. 
    Cases classified as \#Direct are not included in this statistic.

\end{itemize}

\subsubsection{Coverage of Generated Test Inputs}
The system first generates unit test inputs for these sensitive functions.
We leverage \tool{JaCoCo} for Java and \tool{Pytest-cov} for Python to measure line and branch coverage of these test inputs.
On average, the Java test inputs achieve 98.2\% line coverage and 100\% branch coverage, while the Python test inputs achieve 100\% for both line and branch coverage.
The reason Java does not reach 100\% coverage is that when a sensitive function is defined as a static method within a class, our test inputs do not trigger the class’s initialization function.
However, if we only calculate coverage on sensitive functions, we can achieve 100\% coverage.

\subsubsection{Initial Transformation}\label{subsub:eva:initial}
Initially, referring to the first step in Section~\ref{subsec:background:sensitive}, we applied several prompts and examples to transform the code into Rust.
The \textit{\#Direct} row in Table~\ref{tab:eva:consistent} presents the results.
The table shows that, solely with the provided prompts and examples, the LLM agent is unable to successfully transform the majority of the cases, achieving a success rate of less than 21\%.
We manually analyzed these \textit{Direct} cases and summarized their characteristics.
Typically, they contain only a few simple statements, including direct processing of strings (e.g., hashing) and returning the result without any branching.
However, a high failure rate indicates that only providing prompts and examples cannot meet our transformation requirements, as other cases may contain complex operations, such as standard library imports, initialization of cryptographic algorithms, and pertinent error handling.

\begin{table*}[htb]
\centering
\caption{Performance of \system for transforming sensitive functions. }
\label{tab:eva:consistent}
\begin{threeparttable}
\begin{tabular}{l cccc c cccc}
\toprule[1.5pt]
\multirow{2}{*}{\textbf{Metric}} 
  & \multicolumn{4}{c}{\textbf{Java (232 samples)}} 
  & \phantom{a} 
  & \multicolumn{4}{c}{\textbf{Python (153 samples)}} \\
\cmidrule(r){2-5} \cmidrule(l){7-10}
  & \textbf{GPT-4o} & \textbf{DeepSeek} & \textbf{Qwen2.5} & \textbf{LLaMA3.1}
  && \textbf{GPT-4o} & \textbf{DeepSeek} & \textbf{Qwen2.5} & \textbf{LLaMA3.1} \\
\midrule[0.8pt]
\#Direct      & 48 (20.7\%) & 43 (18.5\%) & 41 (17.6\%) & 0 (–) 
              && 29 (18.9\%) & 31 (20.3\%) & 20 (13.1\%) & 0 (–) \\
\#Equivalent  & 213 (91.8\%) & 203 (87.5\%) & 192 (82.7\%) & 4 (1.7\%) 
              && 129 (84.3\%) & 117 (76.5\%) & 103 (67.3\%) & 4 (2.6\%) \\
Ave. Iter.    & 5.3 & 5.5 & 6.5 & 9.5 
              && 5.6 & 5.8 & 6.8 & 13.3 \\
\bottomrule[1.5pt]
\end{tabular}
\end{threeparttable}
\end{table*}

\subsubsection{Iterative Refinement}\label{subsub:eva:iterative}
For remaining cases that did not achieve successful transformation, \system carried out \emph{iterative refinement} with the \tool{ReAct} strategy.
The \textit{\#Equivalent} row in Table~\ref{tab:eva:consistent} illustrates the results that achieve successful transformation.
Meanwhile, the \textit{Ave. Iter.} row represents the average number of iterations required to attain success.
Overall, the success rate of GPT-4o is the highest, regardless of the programming language employed.
In Java transformation tasks, GPT-4o achieved a 91.8\% success rate (213 cases), followed by DeepSeek at 87.5\% (203 cases), and Qwen2.5 at 82.7\% (192 cases).
LLaMA3.1 lagged behind, transforming only 4 cases with a 1.7\% success rate.
Python transformations require more iterations and yield lower overall success rates.
GPT-4o transformed 129 cases with an 84.3\% success rate.
Qwen2.5 managed a 67.3\% success rate on 103 cases, while DeepSeek achieved 76.5\% success on 117 cases.
LLaMA3.1 again lagged significantly, transforming only 4 cases with a 2.6\% success rate.

To assess the statistical differences in performance among GPT-4o, DeepSeek, Qwen2.5, and LLaMA3.1 we conducted pairwise \emph{McNemar's tests}~\cite{mcnemar1947note}. 
For a given set of function instances, this test compares the results of two models on the same instances and evaluates whether their disagreements are systematically biased toward one model. Specifically, it examines instances in which one model successfully transforms a function while the other fails, and tests whether such asymmetries are statistically significant. 
The results show that GPT-4o significantly outperforms DeepSeek, Qwen2.5, and LLaMA3.1 on Java and Python ($p < 0.05$). 
In addition, DeepSeek significantly outperforms Qwen2.5 and LLaMA3.1 on both Java and Python ($p < 0.05$), indicating that it consistently ranks second and, due to its open-source nature, represents a recommended base LLM for local deployment in privacy-critical scenarios.

\subsubsection{Category Influence}\label{subsub:eva:category}
Additionally, Table~\ref{tab:eva:type} summarizes the success rates of different function categories.
The results indicate that, on average, models perform better on cryptographic operations than on serialization tasks. Specifically, cryptographic operations exhibit average model success rates of 87.3\% for Java and 84.1\% for Python.
In contrast, models achieve somewhat lower but still robust average success rates of 78.6\% in Java and 74.2\% in Python on serialization-related functions.
This performance gap is attributed to the nature of the tasks: cryptographic operations often follow standardized patterns and rely on well-documented libraries (e.g., \func{javax.crypto}, \func{hashlib}), making them easier for models to recognize. In contrast, serialization logic tends to involve more heterogeneous structures, nested data handling, and format-specific transformations, which increase semantic complexity.

\begin{table}[ht]
    \centering
    \caption{Success rates for aggregated function types.}
    \begin{threeparttable}
    \begin{tabular}{l|cccccc}
    \toprule[1.5pt]
    \multirow{2}{*}{\textbf{Type}} 
    & \multicolumn{2}{c}{\textbf{GPT-4o}} 
    & \multicolumn{2}{c}{\textbf{DeepSeek}} 
        & \multicolumn{2}{c}{\textbf{Qwen2.5}} \\
    \cmidrule(lr){2-3} \cmidrule(lr){4-5} \cmidrule(lr){6-7}
    & \textbf{J} & \textbf{P} 
    & \textbf{J} & \textbf{P} 
    & \textbf{J} & \textbf{P} \\
    \midrule[0.8pt]

    Crypto
    & 93.0\% & 86.6\% 
    & 88.2\% & 74.1\% 
     & 82.9\% & 69.6\% \\
    Serial
    & 86.7\% & 80.0\% 
    & 84.4\% & 67.5\% 
    & 82.2\% & 62.5\% \\
    
    \bottomrule[1.5pt]
    \end{tabular}
    \begin{tablenotes}
        \footnotesize
        \item[*] \textbf{J} indicates Java; 
        \textbf{P} indicates Python.
        Due to very few successful samples, results for \textbf{LLaMA3.1} are omitted.
    \end{tablenotes}
    \end{threeparttable}
    \label{tab:eva:type}
\end{table}

\subsubsection{Code Language Influence}\label{subsub:eva:language}
Part of Python's lower success rate compared to Java is attributed to its highly integrated language design and dynamic typing system.
On the one hand, some libraries available in Python offer high-level components that are not directly available in Rust.
For example, in one python example, the \func{Crypto.Util.number} module provides functions such as \func{bytes\_to\_long}, \func{getPrime}, and \func{inverse}.
In contrast, Rust's equivalent types, such as BigUint, only include an implementation of \func{bytes\_to\_long}, while functions like getPrime and \func{inverse} must be generated by the LLM, increasing transformation complexity.
On the other hand, as an interpreted language, Python lacks explicit type declarations, which limits the amount of semantic information available to the LLM. Conversely, the target language, Rust, being statically typed, requires more type information for transformation. 
For verification, we added type annotations for the arguments and return values of failed cases for re-transformation.
This process is done semi-automatically: first, we use \tool{Copilot}~\cite{MicrosoftCopilotRelativity} in VSCode to provide a recommendation; then, we manually assess the recommendation. We manually insert the type annotations and verify that the code still passes the previously generated test inputs.
Table~\ref{tab:eva:add_type} illustrates their results.
Among these cases, GPT-4o successfully transformed an additional 4, Qwen2.5 transformed 10 cases, and DeepSeek transformed 9 cases.
This resulted in success rates of 86.2\%, 73.9\%, and 82.3\%, respectively.
Given that GPT-4o already achieved the highest success rate, its improvement is not substantial.
The success rate of Qwen2.5 has improved most significantly.
DeepSeek had nearly matched the performance of GPT-4o.
It is evident that for interpreted languages, the success rate can be enhanced through the manual addition of type annotations.

\begin{table}[htb]
\centering
\caption{Python transformation results with additional type annotations.}
\label{tab:eva:add_type}
\begin{threeparttable}
\begin{tabular}{lccc}
\toprule[1.5pt] 
& \textbf{GPT-4o} &   \textbf{DeepSeek} & \textbf{Qwen2.5} \\
\midrule[0.8pt]
W/O Ann.& 129 (84.3\%) & 117 (76.5\%) & 103 (67.3\%) \\
W Ann. & 132 (+4, 86.2\%) & 126 (+9, 82.3\%)& 113 (+10, 73.9\%)  \\
\bottomrule[1.5pt]
\end{tabular}
\begin{tablenotes}
    \small
    \item[*] \textbf{W/O Ann.} refers to without type annotation;
    \textbf{W Ann.} refers to with type annotation;
\end{tablenotes}
\end{threeparttable}
\end{table}

\subsubsection{Code Complexity Influence}\label{subsub:eva:complexity}
To investigate how code complexity affects the accuracy of sensitive code identification,
we analyzed model performance with respect to the number of LoC in each function, as a practical measure of code complexity.
We categorized all functions into tertiles three roughly equal-sized groups) based on their LoC distribution for each programming language. 
This ensures that our analysis of performance across different function complexities is based on a balanced number of samples in each category.
This resulted in specific boundaries: for Java, the three groups correspond to $\leq 10$, $11-25$, and $>25$ LoC, which contains 77, 78 and,77 samples respectively; for Python, the three groups correspond to $\leq 9$, $10-21$, $>21$ LoC, where each group contains 51 samples.

The results are summarized in Table~\ref{tab:eva:success_size}. 
For both Java and Python, all models achieved stable performance across the categories, with success rate variations within ±2.5\%. These results indicate that the performance remains more or less stable with the increased complexity.

\begin{table}[htb]
\centering
\caption{The success rate under different function sizes.}
\label{tab:eva:success_size}
\begin{tabular}{lcccc}
\toprule[1.5pt]
\textbf{Language} & \textbf{LoC} & \textbf{GPT-4o}  & \textbf{DeepSeek} & \textbf{Qwen2.5}\\
\midrule
& $\le10$ & 91.4  &  88.2 &  82.8 \\
Java     & $11\sim25$ &  91.3  &  87.5  &   82.7\\
     & $>25$ &  91.4  &  85.7 &  80.0\\
\midrule[0.8pt]
 & $\le9$ & 84.1 &  76.8 &  66.7  \\
Python       & $10\sim21$ &  85.2  &  77.0 &  67.2 \\
         & $>21$ &  82.6 &  73.9 &  65.2  \\
\bottomrule[1.5pt]
\end{tabular}
\end{table}

\subsubsection{Integration and System-Level Testing}
To assess functional equivalence, we leveraged each project's existing unit test suite. The validation process involved replacing the original Java/Python function bodies with proxies that preserved the original signatures but forwarded inputs to the TEE-hosted Rust implementation and returned its outputs. We then executed the full unit test suite for each project to confirm that the number of test failures did not increase compared to the original, unmodified project. 
We used the transformations generated by GPT-4o for this experiment, as they consistently achieved the highest success rate in our other evaluations. 
The scope was restricted to projects with unit test suites executable via standard build tools (e.g., Maven, Gradle, pytest), excluding those requiring specialized runtimes like the Android SDK or a GUI.
The resulting benchmark comprises 21 Java projects and 6 Python projects, encompassing 1,248 unit tests for Java and 328 for Python, respectively. 
The Java projects contained 94 of our transformations, while the Python projects included 15 transformations. 
Our baseline execution of the original tests resulted in 12 errors and 22 skips for Java, and 5 errors and 6 skips for Python. 
Our validation criterion was that no new tests failed after integrating the transformed Rust code.

The results showed that for Java, two of the 94 transformations led to new test errors. 
For Python, one of the 15 transformations introduced a failure. 
Manual inspection revealed that these failures were not caused by logical errors in the transformed code but by environmental incompatibilities with the TEE. The Java failures were due to an attempt to invoke a dynamic link call for an elliptic curve algorithm, which is unavailable in the TEE. 
The Python failure was caused by reliance on reading a temporary key from a local file, an I/O operation not permitted across the TEE boundary.

\subsubsection{Failure Analysis}\label{subsub:eva:failure}
\label{subsec:eva:failure}
\begin{figure*}[htb]
\centering
\begin{minipage}{.49\textwidth}
\begin{tcolorbox}[colback=gray!5!white, colframe=cyan!40!black, boxsep=2pt, top=1mm, bottom=1mm, left=0pt, right=0pt, title=\centering{Original Python code}]
\begin{lstlisting}[style=CodeStyle]
from passlib.hash import ldap_salted_sha1
def hash(a):
    # hash salt
    h = ldap_salted_sha1.hash(a)
    
    return h
\end{lstlisting}
\end{tcolorbox}
\end{minipage}
\hfill
\begin{minipage}{.49\textwidth}
\centering
\begin{tcolorbox}[colback=gray!5!white, colframe=cyan!40!black, boxsep=2pt, top=1mm, bottom=1mm, left=0pt, right=0pt, title=\centering{Transformed Rust code}]
\begin{lstlisting}[style=CodeStyle]
use sha2::{Sha256, Digest};use hex;
fn hash(a: &str) -> String {
    let mut hasher = Sha256::new();
    hasher.update(a);
    hex::encode(hasher.finalize())
}
\end{lstlisting}
\end{tcolorbox}
\end{minipage}
\caption{Inequivalent functionality caused by security method change.}
\label{code:eva:hashing}
\end{figure*}

For the code that failed to transform, we manually inspected its implementation and structure to identify key issues, excluding LLaMA3.1 from this analysis due to its extremely low success rate.
\begin{description}[leftmargin = 0pt]
    \item [Sophisticated cryptography (46.2\%).]
    If a code employs sophisticated and multiple algorithms, the LLM may fail to accurately transform the code. 
    For instance, Java code simultaneously utilizes \tool{Curve25519} public keys and \tool{XSalsa20} as a stream cipher for encryption, ultimately employing the \tool{Poly1305} algorithm for authentication.
    \system inaccurately transforms the code due to the complexity of the cryptographic operations, preventing successful equivalence validation before the iteration threshold was reached.

    \item [Shift operations (22.9\%).]
    Even if the code transformation appears successful, many transformed programs with bit shift operations terminate with execution errors.
    For these cases, the model translations are wrong. There are two possible reasons: (i) LLMs lack the precise reasoning required for bit-shifting; (ii) there are different semantics of bit shift operations among Java, Python, and Rust. It is hard to determine if only one or both factors influence the wrong translations and which one is the most influential.

    \item [Missing flag variable (14.7\%).] There is a global static flag whose specific value the LLM cannot determine, resulting in continuous incorrect substitution of variables.

    \item [Functionality change (9.1\%).]
    Among the cases that did not pass equivalence validation, there are instances of \dquote{security upgrade} modifications.
    For example, the original code employed \emph{SHA-1} for hash computation, while the transformed code replaced it with \emph{SHA-256}, thereby enhancing security.
    Although this results in bias, from a security standpoint, \emph{SHA-256} offers greater reliability than \emph{SHA-1},
        given that \emph{SHA-1} has been demonstrated to be more vulnerable to collision attacks~\cite{Merrill17limits, wang2005finding}.
    Subjectively, we consider this to be a beneficial modification as it offers a more secure implementation of functionality.
    Unfortunately, the LLM did not always exhibit such reliability.
    In the absence of sufficient semantic information, 
        particularly with Python code, 
        the LLM may compromise the security of the transformed code while also producing different outputs.
    For example, Figure~\ref{code:eva:hashing} is the original and transformed Python code, which calculates the hash value using the hashing and salting method (recursive hash).
    Although the LLM substituted \emph{SHA-1} with \emph{SHA-256}, the omission of the salting mechanism still led to a functional inequivalence.

    \item [Others (7.1\%).] There are no obvious code features; we believe it is a problem with the model's own ability.
\end{description}

The distribution of errors indicates that the model struggles the most with shift operations and sophisticated cryptography. This is mainly because the two tasks require strong capability of mathematical reasoning and bit-level data manipulation, making it a challenge for LLMs to infer the underlying logic from text alone and thus leading to error-prone transformation.

\subsubsection{Comparison and Analysis}\label{subsub:eva:comparison}
\system employs multiple methods to assist the LLM in accomplishing the transformation tasks.
This sub-experiment still employs LLMs for transformation but uses different methods for comparison:
\begin{itemize}[leftmargin = *]
    \item \textbf{Zero-Shot:} This method directly prompted the LLM to transform sensitive functions into Rust code without providing any additional information.
    
    \item \textbf{Few-Shot:} This method prompted the LLM to transform the code using three transformed examples.
    However, unlike the \emph{initial transformation}, it did not require the LLM to analyze the code first.

    \item \textbf{Compilability Check:} Other processes are consistent with \system, except that there is no equivalence validation during the iterative refinement.
    It only utilized compiler checks as feedback for the LLM agent.
\end{itemize}

Table~\ref{tab:eva:cmp} presents the comparison results.
According to the table, the success rate for the \textit{Zero-Shot} is the lowest due to the minimal information provided.
When additional transformation examples were introduced (\textit{Few-Shot}), the success rate improved.
Despite this enhancement, a significant number of cases (over 80\%) still remained unsuccessful.
Following the introduction of the \textit{Compilability Check}, the success rate exhibited a substantial increase.
\textit{Compilability Check} resulted in the majority of the transformed code being executable; however, it did not ensure its equivalence to the original code.
In comparison, since equivalence validation was introduced, \system achieved the highest success rate, ensuring both executable and equivalent functionality of transformed code.

\begin{table*}[htb]
    \centering
    \caption{Comparison of code transformations across different methods.}
    \label{tab:eva:cmp}
    \begin{tabular}{lcccccccc}
    \toprule[1.5pt]
    \multirow{2}{*}{\textbf{Method}} & \multicolumn{4}{c}{\textbf{Java (232 \#Sensitive Samples)}} & \multicolumn{4}{c}{\textbf{Python (153 \#Sensitive Samples)}} \\
    \cmidrule(lr){2-5} \cmidrule(lr){6-9}
    & \textbf{GPT-4o} & \textbf{DeepSeek} & \textbf{Qwen2.5} & \textbf{LLaMA3.1} & \textbf{GPT-4o} & \textbf{DeepSeek} & \textbf{Qwen2.5} & \textbf{LLaMA3.1}\\
    \midrule[0.8pt]
    
    Zero-Shot & 22 (9.4\%) & 20 (8.6\%) & 18 (7.8\%) & 0 (-) & 13 (8.5\%) & 13 (8.5\%) & 11 (7.2\%) & 0 (-) \\
    Few-Shot & 44 (18.9\%) & 42 (18.1\%) & 39 (16.8\%) & 0 (-) & 26 (16.9\%) & 28 (18.3\%) & 18 (11.8\%) & 0 (-) \\
    Compilability Check & 158 (68.1\%) & 148 (63.8\%) & 140 (60.3\%) & 2 (0.8\%) & 61 (39.8\%) & 57 (37.2\%) & 49 (32.1\%) & 1 (0.6\%) \\
    \system & 213 (91.8\%) & 203 (87.5\%) & 192 (82.7\%) & 4 (1.7\%) & 129 (84.3\%) & 117 (76.5\%) & 103 (67.3\%) & 4 (2.6\%)\\

    \bottomrule[1.5pt]
    
    \end{tabular}
    \end{table*}

\subsubsection{Stability}\label{subsub:eva:stability}
Utilizing LLMs as agents introduces an element of randomness, which can lead to different transformation outcomes for the same code.
Thus, we also evaluate the output stability of \system. To reduce API token costs, we randomly sampled approximately 10\% of the cases from the original Java and Python datasets for a stability assessment. 
The final stability assessment dataset comprised 37 instances (i.e., sensitive functions), of which 23 instances came from the Java dataset and 14 instances from the Python dataset. 

We conducted four more rounds of repeated experiments, and the results are presented in Table~\ref{tab:eva:rep}.
For these 37 samples, GPT-4o, Qwen2.5, and Deepseek successfully transformed 35, 30, and 33 samples, respectively, in the original experiment.
The table indicates that, in most cases, the success rate remained equivalent.
However, the results also exhibited minor fluctuations.
In \textit{Exp1}, GPT-4o successfully transformed one additional sample, whereas Qwen2.5 in \textit{Exp3} failed to transform one previously successful sample.
The additional successful transformation was a data encryption process.
The failure instance arose from a functionality change, where the source code used \emph{SHA1}, while the transformed result applied \emph{SHA256}.
In general, despite the presence of some fluctuations, \system demonstrates stability in its transformation outcomes.

\begin{table}[ht]
    \centering
    \caption{Success rates of transformations in repeated experiments.}
    \label{tab:eva:rep}
    \begin{threeparttable}
    \begin{tabular}{l|c|cccc}
    \toprule[1.5pt]
    & \makecell[c]{\textbf{Original Exp} \\ \textbf{(37 Samples)}} & \textbf{Exp1} & \textbf{Exp2} & \textbf{Exp3} & \textbf{Exp4} \\
    \midrule[0.8pt]
    GPT-4o & 35 Equivalent & 36 (+1) & 35 & 35 & 35 \\
    DeepSeek & 33 Equivalent & 33 & 33 & 33 & 33 \\
    Qwen2.5 & 30 Equivalent & 30 & 30 & 29 (-1) & 30 \\
    \bottomrule[1.5pt]
    \end{tabular}
    \begin{tablenotes}
    \small
    \item[*] \textbf{Exp} refers to Experiment.
    \end{tablenotes}
    \end{threeparttable}
\end{table}

\subsection{RQ3: Resource Consumption}\label{subsec:eva:resource}
This experiment exhibits the resource consumption of \system, including LLM consumption and practical TEE consumption.

\subsubsection{Resource Consumption in LLMs}\label{subsub:eva:llm}

This study compares the token consumption of LLMs for successful transformations in our experiments.
Table~\ref{tab:time_consumption} reports the average token usage and monetary cost across the Identification (Ident.), Test Generation (Test Gen.), and Transformation (Trans.) stages, as well as the average cost per iteration incurred during transformation phases (Per. x \#Iter). 
From the table, we observe the following: (1) The token usage of various models during the sensitive function identification phase is similar (approximately 0.48K). 
(2) The primary differences in token consumption occur in the transformation phase. LLaMA3.1 requires more iterations in the transformation phases, resulting in higher token usage. The other three models perform better than LLaMA3.1. 
(3) GPT-4o and DeepSeek generate more detailed reasoning within the ReAct iterative process, leading to higher per-iteration token usage but fewer overall iterations. 
(4) DeepSeek is cheaper than GPT-4o. While its total token consumption is comparable to GPT-4o, its cost is lower (\$0.024 vs. \$0.398), providing a cheaper choice. Qwen2.5 and LLaMA3.1 are deployed locally, so they incur no API costs.

\begin{table}[htb]
\centering
\caption{Average token usage and cost of LLMs.}
\label{tab:time_consumption}
\begin{threeparttable}
\begin{tabular}{lcccc}
\toprule[1.5pt]
\multirow{2}{*}{\textbf{Model}} &  \multicolumn{3}{c}{\textbf{Avg. Token Usage (K)}} & \multirow{2}{*}{\textbf{Avg. Cost}} \\
\cmidrule[0.8pt](lr){2-4}
 & {\textbf{Ident.}} & {\textbf{Test Gen.}} & {\textbf{Trans.(Per. x \#Iter)}} & \\
\midrule
GPT-4o     & 0.49K & 12.4K & 36.72K (6.8K x 5.4) & \$0.398 \\
DeepSeek & 0.48K & 13.2K &  39.1K (6.9K x 5.6) & \$0.024 \\
Qwen2.5  & 0.48K & 13.6K &  40.4K (6.1K x 6.6) & -- \\
LLaMA3.1 & 0.49K & 17.3K & 65.9K (5.7K x 11.4) & -- \\
\bottomrule[1.5pt]
\end{tabular}
\begin{tablenotes}[para,flushleft]
\footnotesize
\item \textbf{Ident.}: Average token usage during the identification stage; \textbf{Test Gen.}: Average token usage during the test generation stage; \textbf{Trans.}: Average token usage of the successful transformation cases. 
Values in parentheses denote average tokens per iteration x average number of iterations.
Qwen2.5 and LLaMA3.1 are locally deployed and incur no cost.
\end{tablenotes}
\end{threeparttable}
\end{table}

\subsubsection{Practical Resource Consumption of Execution}\label{subsub:eva:execution}
We present several case studies to illustrate the practical resource consumption of code executed within the SGX TEE.
We used the \tool{Rust-SGX}~\cite{rustsgx} tool to compile, sign, and load the transformed code into the TEE.
For experimental convenience, we implemented a basic attestation service inside the SGX.
This service accepts a quote, verifies whether it was generated by the expected TEE instance, and checks whether the content has been altered.

Table~\ref{tab:time_run} compares the resource consumption of the original code with that of the transformed code executed inside the TEE.
The table includes the following columns: whether an external library is used (Lib.), the original execution time before transformation (Ori. Cost), the execution time inside the TEE (TEE Cost), and the performance overhead expressed as a multiplicative increase (Increase).
Each cost value represents the average of five runs to ensure measurement stability.
The measurements account for TLS communication overhead, quote verification cost, and function execution within the TEE.

In general, the resource consumption of most code executed inside the TEE increases due to transmission delays and attestation verification.
Among the evaluated cases, Python's \func{SHA256} implementation relies on bitwise shift operations and does not use external libraries, resulting in a relatively low execution time in its original version (0.2 ms) but experiencing a significant increase (251 ms) when executed inside the TEE.
Similarly, \func{AES-CFB Enc.}, which utilizes a C-based library, also has a low original execution time and exhibits an overhead of approximately 143$\times$ when ported into the TEE. 
Conversely, for computationally intensive functions like \func{RSA Key Gen.}, the original execution time (212 ms) is already substantial. 
In this case, the fixed TEE overhead constitutes a smaller fraction of the total runtime, resulting in a much lower relative increase of only 1.6$\times$.

\begin{table}[htb]
\centering
\caption{Runtime overhead of transformed functions executed inside the TEE.}
\label{tab:time_run}
\begin{threeparttable}
\begin{tabular}{lcccc}
\toprule[1.5pt]
\textbf{Case} & \textbf{Lib.} & \textbf{Ori.\ Cost} & \textbf{TEE Cost} & \textbf{Multiple} \\
\midrule[0.8pt]
MD5 (Java)            & \ding{51} & 28ms    & 338ms   & 12.1$\times$ \\
SHA1 (Java)           & \ding{53} & 18ms    & 253ms   & 13.1$\times$ \\
Random (Java)         & \ding{51} & 18ms    & 254ms   & 14.1$\times$ \\
AES Decryption (Java)    & \ding{51} & 42ms    & 242ms   & 5.7$\times$  \\
Serialization (Java)  & \ding{53} & 13ms    & 253ms   & 19.4$\times$  \\
RSA Key Gen.\ (Python)    & \ding{51} & 212ms   & 355ms   & 1.6$\times$  \\
SHA256 (Python)       & \ding{53} & 0.2ms   & 251ms   & 1255$\times$ \\
AES-CFB Enc. (Python)  & \ding{51} & 1.2ms   & 172ms   & 143$\times$ \\
\bottomrule[1.5pt]
\end{tabular}
\begin{tablenotes}
    \small
    \item[*] \textbf{Lib.} indicates whether an external library is used in original code;
    \textbf{Ori.\ Cost} refers to the original execution time before transformation; \textbf{Key Gen.} is short for key generation; 
    \textbf{Enc.} is short for encryption
\end{tablenotes}
\end{threeparttable}
\end{table}

\section{Discussion}
\label{sec:discussion}
\system's transformations follow the standard TEE application migration pattern~\cite{lind2017glamdring}, which separates the application into a trusted part and an untrusted part and establishes a secure communication channel.
\system preserves integrity and confidentiality through a multi-layered security design. 
To ensure integrity, \system integrates standard remote attestation to verify code authenticity before execution and transforms sensitive functions into memory-safe Rust, effectively mitigating runtime memory corruption vulnerabilities (e.g., buffer overflow). 
To ensure confidentiality, the \emph{Execution Linker} of \system establishes a secure TLS channel for all cross-boundary interactions, mandating that all function arguments are serialized and encrypted during transit to prevent data leakage at the interface level.

\system reduces the engineering burden of adopting TEEs by automating phases that normally are done manually. 
First, in partitioning, unlike manual methods that need labor-intensive dependency analysis to establish trust boundaries, \system automatically identifies and protects sensitive functions, which avoids the difficulty of untangling coupled application logic.
Second, \system replaces the manual, error-prone process of porting high-level code into Rust with an LLM-driven transformation pipeline. This transformation eliminates the need for TEE-specific systems expertise during the application development lifecycle. 
Finally, for validation, \system reduces the high costs of manual testing. By combining the ReAct strategy with automated differential testing, \system checks functional equivalence between the original and trusted components, freeing developers from writing and maintaining extensive test suites.

\section{Threats to Validity and Limitation}
\label{sec:validity}
\subsection{Threats to Validity}
\vspace{0.1cm}
\noindent    
{Internal Validity.}
\rain{In \system, we assess the equivalence of code transformations by executing the original and transformed functions on the same set of inputs and comparing their outputs. We acknowledge that the current validation may not capture all semantic aspects. To mitigate this threat, we use LLMs to generate test inputs with high coverage, improving test adequacy by maximizing line and branch coverage. 
Because \system targets leaf functions with simple control flow and limited dependencies, the generated tests achieve 100\% line and branch coverage in our experiments. This high coverage ensures that diverse execution paths are exercised and increases the likelihood that behavioral divergences, if present, are exposed, thereby strengthening confidence in the equivalence of the tested functions.}

\vspace{0.1cm}
\noindent    
{External Validity.}
A potential threat is that our results may not generalize to other models or datasets. To mitigate this threat, we evaluate four representative LLMs on 385 labeled sensitive functions spanning multiple categories. This evaluation across diverse models and a nontrivial number of functions helps reduce model-specific or dataset-specific bias. 
Nevertheless, our study does not cover all possible models or application scenarios. We therefore encourage future work to extend this evaluation to additional models, datasets, and domains.

\vspace{0.1cm}
\noindent    
{Construct Validity.}
One potential threat arises from the operationalization of \dquote{sensitive functions}. 
We acknowledge that the identification of sensitive functions may be subject to bias or omission. 
To mitigate this threat, we provide explicit labeling criteria and guidelines. 
As described in Section~\ref{subsec:background:sensitive}, we limit the scope of sensitive functions to those well-defined categories.
This restricted scope reduces subjectivity by grounding the construct in established literature. 
All functions are labeled according to these predefined criteria, and the process is conducted independently by multiple authors with relevant expertise, with disagreements resolved through discussion. These measures improve the objectivity and reproducibility of the labeling process.

An additional threat to construct validity lies in our dataset construction process.
Due to the large number of functions labeled (10,984 functions), the process is rather adhoc.
Thus, to ensure the quality of the dataset, we performed a second round of labeling with statistically significant, randomly selected samples (372, 95\% confidence level with a 5\% margin of error). 
This second round of labeling followed a more rigorous protocol. 
As described in Section~\ref{subsec:background:sensitive}, we restricted the definition of sensitive functions to well-established categories informed by prior TEE-related studies~\cite{9276587,paju2023sok}, including \emph{Encryption}, \emph{Decryption}, \emph{Signature}, \emph{Verification}, \emph{Hash}, \emph{Seed Generation}, \emph{Random Number Generation}, \emph{Serialization}, and \emph{Deserialization}. 
For each function, we applied the following ordered guidelines to determine sensitivity: (1) checking for explicit comments indicating the presence of sensitive operations; (2) identifying invocations of known cryptographic or deserialization APIs (e.g., \tool{hashlib.sha256}, \tool{cryptography.fernet.Fernet}); and (3) determining whether the function implements a complete sensitive logic pattern, such as AES-based encryption. 
We employed a quality-control process in which two authors independently reviewed each sample, while a third author adjudicated any disagreements. 
Each function in the sampled subset was independently reviewed by two authors. The inter-rater agreement reached an \dquote{almost perfect} level~\cite{viera2005understanding}, with a Cohen’s $\kappa$ of 0.907. 
The remaining 11 disagreements were resolved through adjudication by a third author. We find that the labeling outcomes for the sampled subset are consistent with the initial labeling results in 97.5\% of cases, providing strong evidence for the reliability of the dataset.

\subsection{Limitation}
\system currently hardcodes environment variables in the transformed code.
This approach may not be suitable for migrating our generated code across different hardware platforms.
To deal with this issue, humans can manually check those hardcoded environment variables and modify them. 
Our goal is to automate some steps but not to fully exclude humans in the loop.

For experiment, our dataset is primarily composed of cryptographic projects, which may suggest limited diversity in software domains. We intentionally selected these projects to ensure a sufficient number of sensitive functions, which serve as positive samples in our experiments. 
Although these projects center on cryptography, they have diverse application contexts like Android key management applications, and cloud service plugins. 
We acknowledge that, in many real-world workloads, the proportion of sensitive functions is typically lower than in our dataset. However, the methodology of \system is domain-agnostic and is not restricted to cryptographic projects. As a result, the techniques proposed in \system remain applicable to conventional software projects. 
In future work, we plan to broaden our evaluation to include a wider range of projects to further assess generalizability.

\section{Related Work}
\label{sec:related}
\subsection{Application Porting to TEE}\label{subsec:related:porting}
\tool{Lejacon}~\cite{miao2023lejacon} developed a customized JVM approach to run Java applications on SGX, isolating them within a secure environment to protect the confidentiality of code and data. While successful, this method increases the Trusted Computing Base (TCB) size, expanding the attack surface and potential security risks.
\tool{TEESlice}~\cite{zhang2024no} enhances the security of on-device machine learning by accurately isolating privacy-sensitive weights in models. It employs a partition-before-training strategy, dividing models into backbone and private slices, and uses a dynamic pruning algorithm to optimize slice size, ensuring that they fit within TEEs without loss of accuracy.
\tool{MyTEE}~\cite{han2023mytee} provides TEE functionality for embedded devices lacking full TrustZone support. It utilizes stage-2 page tables for TEE isolation from the untrusted OS and implements a DMA filter to prevent malicious memory access. Secure I/O is achieved by delegating peripheral requests to the untrusted OS, while essential components are protected by paging.
For safeguarding confidential environments, \tool{ACAI}~\cite{Supraja294554} utilizes Arm's Confidential Computing Architecture (CCA) to securely integrate accelerators like GPUs and FPGAs. This method extends CCA's security features to these devices, ensuring isolation at the virtual machine level and protection against both software and physical threats.
To address the excessively large Trusted Computing Base in TEE application adaptation, Lind et al.~\cite{lind2017glamdring} propose a partitioning method that extracts security-sensitive components and places them within a secure enclave, ensuring that the enclave's code does not compromise data integrity or confidentiality.

In contrast, we migrate only a small portion of critical code into the TEE as native code. This approach allows \system to maintain functionality while avoiding the introduction of additional runtime overhead.

\subsection{Code Transformation}\label{subsec:related:transformation}
Hong et al.~\cite{hong2023improving} propose enhancing C-to-Rust translation through the automatic replacement of unsafe features with safe alternatives.
While Rust's ownership type system prevents memory and thread bugs, current translators only perform syntactic conversions, requiring developers to perform manual refactoring.
Their approach focuses on replacing unsafe features, such as the lock API and output parameters.
However, their approach is limited to C transformations.
\tool{CodeStylist}~\cite{ting2023codestylist} utilizes neural methods for code style transfer, building upon existing code language models pre-trained on extensive open-source codebases and fine-tuned through a multi-task training approach to handle diverse style transformations.
However, it requires a substantial amount of pre-training data.
\tool{BabelTower}~\cite{wen2022babeltower} presents a learning-based framework that automates the transformation of sequential C code into parallel CUDA code. It effectively addresses GPU programming complexities by employing an extensive dataset of compute-intensive functions and utilizing back-translation, combined with a discriminative reranker, to handle unpaired corpora and semantic transformations.
\tool{hmCodeTrans}~\cite{liu2024hmcodetrans} introduces a method for interactive human-machine collaboration in code translation. The approach defines two collaboration patterns: prefix-based and segment-based, which allow software engineers' edits to guide the model toward improved retranslation.
Furthermore, response time is reduced through an attention cache module that avoids redundant prefix inference and a suffix splicing module that minimizes unnecessary suffix computation.

We utilize an LLM to assist in code transformation, while simultaneously incorporating compilability checks and equivalence validation to ensure the functional equivalence of the transformation.
This process does not require prior training data.

\section{Conclusion and Future Work}
\label{sec:conclusion}
Despite the strong security protections offered by TEEs, adapting existing programs to leverage these guarantees remains challenging due to the need for extensive domain knowledge and manual intervention. 
The limited resources of TEEs and the complexity of identifying and transforming security-sensitive functions have hindered their broader adoption. To address these challenges, we propose \system, an automated LLM-enabled approach. 
\system automatically identifies, partitions, and transforms sensitive functions in high-level language programs (e.g., Java and Python), porting them to TEEs while ensuring compatibility with secure environments like Intel SGX through secure communication channels.
Specifically, \system incorporates robust iterative transformation and validation mechanisms, utilizing compiler feedback for compilability and functional equivalence checks to guarantee the executability and functional consistency of the TEE-adapted code. 
Our evaluation,  when applying GPT-4o for transformation, \system achieved high success rates in porting, with 91.8\% of Java code and 84.3\% of Python code successfully transformed while preserving functional equivalence and successful compilation.

\system is designed with modularity and flexibility in mind, making it adaptable to other types of sensitive code and to additional programming languages. 
Our LLM-driven approach enables source-level analysis and transformation without reliance on specific runtime environments, thus supporting portability across languages provided appropriate parsing and code analysis tools are available. 
\system can be readily extended to sensitive operations, such as network communication, privilege management, or resource-intensive workflows, with recognizable code patterns.

Currently, the migration of leaf functions into the TEE environment is limited to basic external library calls that have direct equivalents in the target TEE. 
In future work, we will broaden this support to cover more intricate invocation scenarios, such as those involving complex data structures and non-standard external libraries.
To strengthen functional equivalence verification, we plan to adopt a more rigorous validation regime that incorporates comprehensive test suites going beyond basic input-based tests, thereby improving the transformation success rate.

The observed integration failures also highlight an area for improvement. 
In our current study, we only use LLM to generate test inputs. In the future, we plan to (i) use LLM to generate integration and system tests, and (ii) use the outcomes of these tests as feedback for \system to refine the transformed program. 
We also plan to develop a solution that can proactively identify functions whose transformation can cause integration errors and exclude those. 

\section*{Acknowledgement}
This research is supported by the National Research Foundation, Singapore, and the Cyber Security Agency of Singapore under its National Cybersecurity R\&D Programme (Proposal ID: NCR25-DeSCEmT-SMU). Any opinions, findings and conclusions or recommendations expressed in this material are those of the author(s) and do not reflect the views of the National Research Foundation, Singapore, and the Cyber Security Agency of Singapore.

\printbibliography

\end{document}